\documentclass[twocolumn,a4]{aastex63}

\usepackage[LGR,T1]{fontenc}	

\usepackage[utf8]{inputenc}

\usepackage{xcolor}			

\usepackage{graphicx}
\usepackage{txfonts}

\usepackage{natbib}
\bibpunct{(}{)}{;}{a}{}{,} 		

\usepackage{enumitem}

\mathchardef\mhyphen="2D

\definecolor{my_orange}{HTML}{ff9500}	
\pagecolor{my_orange}

\definecolor{xlinkcolor}{cmyk}{0,0,1,0}
\definecolor{xlinkcolor}{HTML}{06ff02}

\newcommand{\textgreek}[1]{\begingroup\fontencoding{LGR}\selectfont#1\endgroup}

\graphicspath{{./}{figures/}}

\shortauthors{Through the decades}
\shorttitle{Fundamentally so}
\received{on April 1st, 2023}

\begin{document}

\title{\LARGE The most fundamental question of all times}

\author{Ste Berta}
\affiliation{My Place, Pianeta Terra, 00003 Sistema Solare, Via Lattea. E-mail: ste\_atreb yahoo it}

\author{Avril de Poisson}
\affiliation{Her Place, Plan\`ete Terre, 00003 Syst\`eme Solaire, Voie Lact\'ee.}

\author{Kriemhild von Scherz}
\affiliation{Their Place, Planet Erde, 00003 Sonnensystem, Milchstra{\ss}e.}

\author{Saul Fools}
\affiliation{Some place, Planet Earth, 00003 Solar System, Milky Way.}


\begin{abstract}

In the last few decades, reading the literature, we realized that we Astronomers have a 
strong preference to undertake very ambitious projects, and search for answers to the most fundamental questions in the history of the entire Universe. 
After running multiple times into such cardinal quest, the curiosity became no more sustainable and we had to find out. To our greater surprise, in the last few decades we had been restlessly participating to this superhuman endevour.
Therefore we hereby explore the roots and grounds of this fundamental search, through the past decades, centuries and millennia.


\end{abstract}


\section{Introduction}

On one unspecified day of this century, while reading two completely independent publications in preparation, we stumbled into a rather entertaining discovery. The introductions of the two manuscripts began with the exact same fundamental, decadal sentence: {\it in the last few decades we tried to answer one of the most fundamental questions of astrophysics}, or alike. 

To greater surprise, a deeper investigation of the origin of such ambitious plan highlighted that several other publications of which we were co-authors indeed started with that sentence. And proposals, telescope time projects, funding requests, etc. often aimed at a similar, highly purpose too. 

This mind-blowing discovery triggered a sequence of different emotions in our souls: first entertainment; then amusement and laughter; and finally preoccupation. Therefore came the idea of this fundamental research. After all, we always wanted to give our contribution to the long list of April Fools' ArXiv entries: what better occasion than this? 

Making use of the famous Astrophysics Data System (ADS\footnote{\url{https://ui.adsabs.harvard.edu/classic-form/}}), this fundamental paper studies the incidence of few very popular sentences in the abstracts of the recent and not-so-recent literature. 

Although very effective as ``show opener'' or ``ice breakers'', especially when we are affected by the so-called ``white page syndrome'', our feeling a posteriori is that such frequently used sentences are regrettably empty and do not add much information to the flow of the manuscript. As for proposals, despite being a standard start, these sentences tend to waste the time of referees who have only few minutes to read our project out of hundred others.

Obviously we have no ambition of being complete here: many popular sentences and combinations of words exist. We invite our readers to identify them in their favorite branch of astronomy. We believe that the ``{\it fundamental question}'' and the ''{\it last decade}'' are rather universal and therefore they are our choice of reference.

\begin{figure*}[!ht]
\centering
\includegraphics[width=0.72\textwidth]{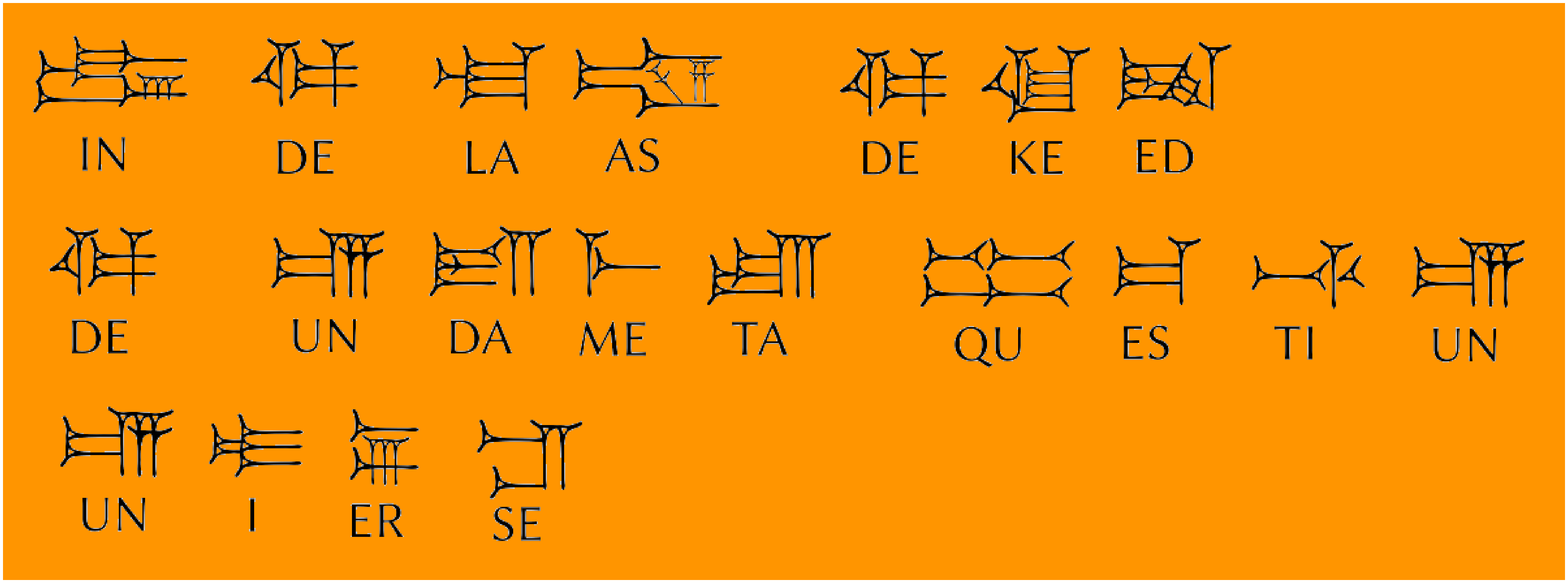}
\caption{Reproduction of the content of clay tablet nr. 654321 from the Royal Library of Ashurbanipal in Niniveh, including translation of cuneiform symbols.}
\label{fig:niniveh}
\end{figure*}

\begin{figure*}[!ht]
\centering
\includegraphics[width=0.72\textwidth]{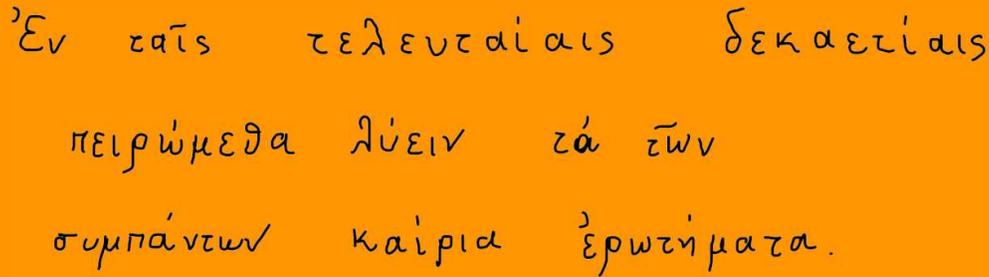}
\caption{Original text by Spyridon from Thebes, 4th centurey B.C.}
\label{fig:greek}
\end{figure*}

This study is limited to Abstracts because searching in the Introductions of papers would imply a more complex process: as a matter of fact we are serious people and we have not much time to waste on this. It is expected that these sentences are much more widely used in Introductions and the numbers presented here would need to be increased by at least a factor of ten if they were taken into consideration. 

Our main goals are to: {\em a)} find out in which decade this all started; {\em b)} which is the most effective (or most used) way to convey to the reader how important and excruciating our quest for the most fundamental answer is; {\em c)} obviously find the ultimate answer.
This manuscript is structured as follows: Section 2 invents a completely fake history of {\it the} fundamental question  through ancient history; Section 3 describes our search in the modern literature and the available data; Section 4 presents our fundamental results; and finally Section 5 draws the conclusions of the last decade and designs some prospects for the next.



\section{Some (fake) ancient history}

In the last few decades, astronomers have tried (in vain) to answer the most fundamental questions that the Universe and Nature posed. As a matter of fact this desperate search for the Absolute Truth can be traced all the way back to the start of the experimental method \citep{galilei1610,galilei1632} through the last few centuries, if not to the ancient greeks philosophers and even to the mathematicians of Babylon through the last few millennia. 

In the Niniveh Royal Library of Ashurbanipal, the clay tablet nr. 654321 reads ``{\it in de la-as deke-ed (...) de undameta questiun unierse}''. Despite the ancient terminology, which might also include some dialectal inflection of the scribe, the meaning is sound and clear. It is thrilling to feel the vibrant emotion of our erudite distant ancestors, transmitted to us by these words carved in time \citep[][see Fig. \ref{fig:niniveh}]{niniveh65432}.

The greek philosopher Spyridon from Thebes followed the path of his ancient Babylon neighbours. In one fragment of his most famous masterpiece he reports that  {\it \textgreek{τις τελευταίες δεκαετίες προσπαθούμε να λύσουμε τα θεμελιώδη ερωτήματα του Σύμπαντος}}, as transliterated to modern greek \citep{spyridon69}. Figure \ref{fig:greek} shows a reproduction of the original ancient text. 

The good old Romans, although mostly interested in conquering the World, were not insensitive to the great questions of the Universe, and the poet Aprilis Fatuus dedicated their best years to this quest. In their memories, we read ``{\it per decennia responsa quaerebamus quaestioni fundamentali universi, nova scientia quaerimus}'' \citep[for decades we searched for answers to the fundamental questions of the Universe, we seek new knowledge,][]{aprilis165}.

Surely these questions were subject of research in all ancient cultures. Old manuscripts found all over the globe testify the efforts of the past civilizations in this fundamental search. Although very interesting per s\'e, a thorough report of all known ancestral evidence goes beyond our scope. Therefore now it is time to come back to post-Galilean modern astronomy.


\begin{figure*}[!ht]
\centering
\includegraphics[width=0.43\textwidth]{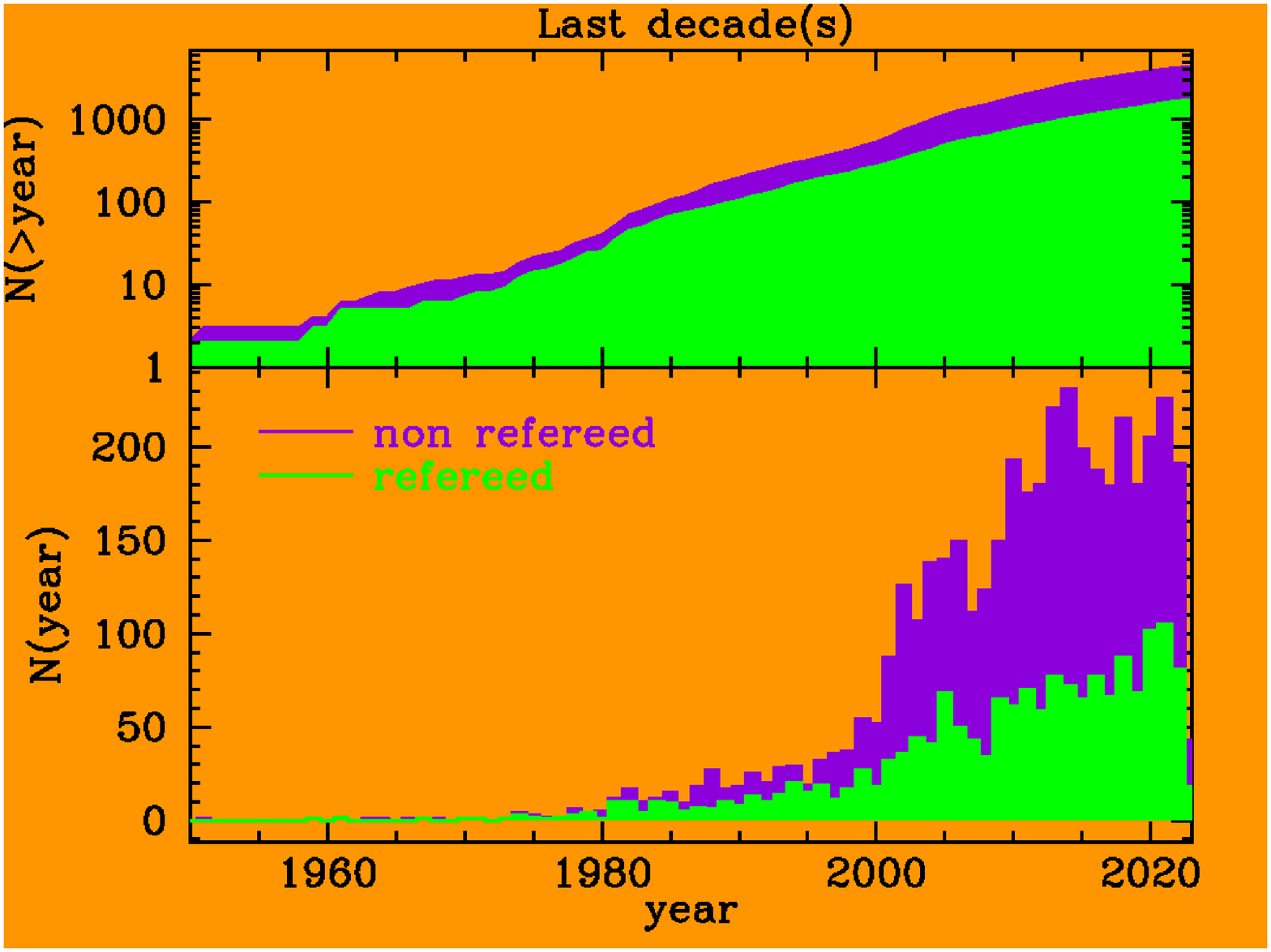}\\
\includegraphics[width=0.43\textwidth]{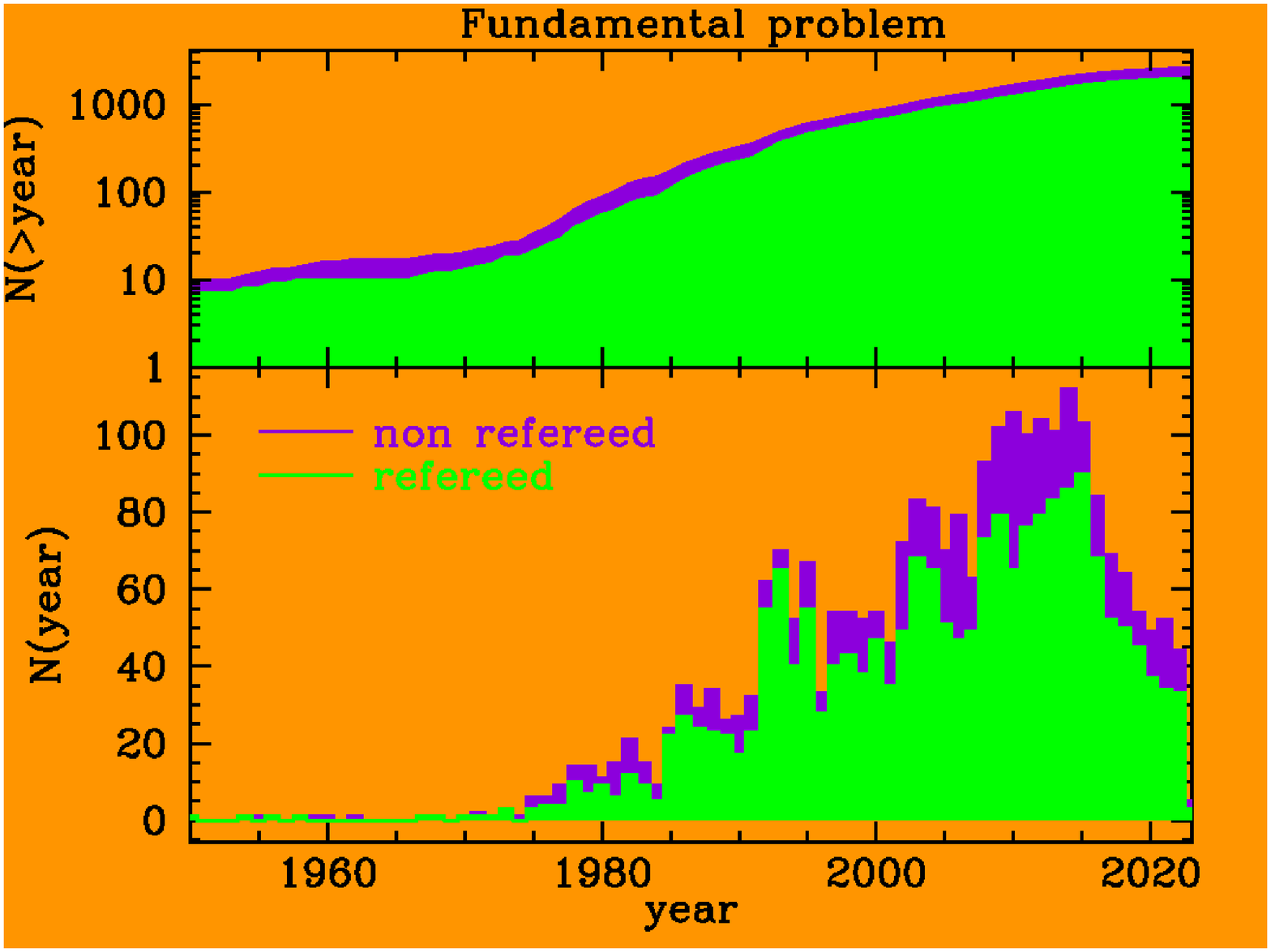}
\includegraphics[width=0.43\textwidth]{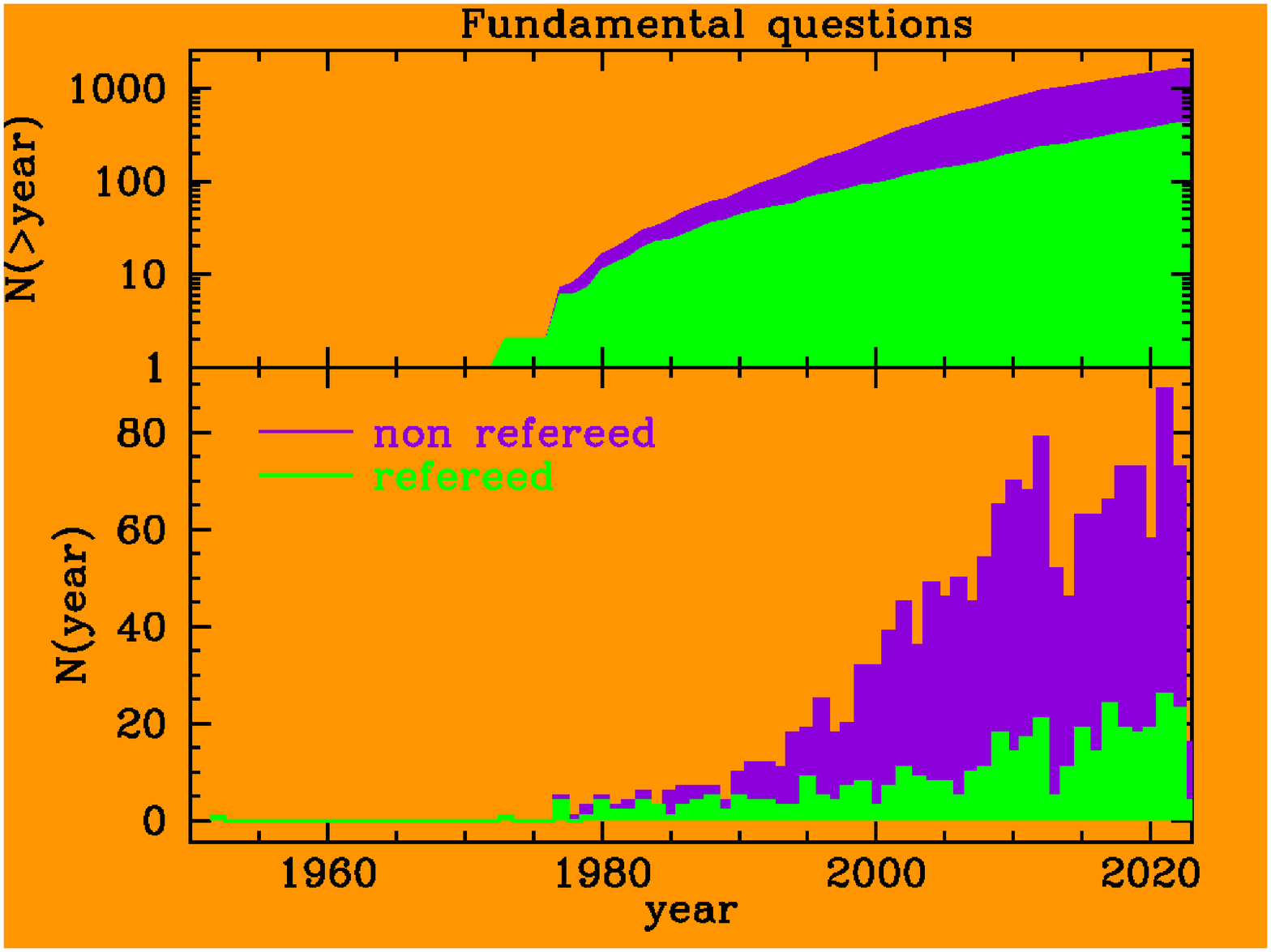}
\caption{Distribution of the most famous expressions in astronomy abstracts since the year 1950.}
\label{fig:metrics}
\end{figure*}

\begin{table*}[!ht]
\centering
\begin{tabular}{lcccc}
\hline
\hline
Expression & Years range & Number of entries & Refereed & h-index\\
\hline
Last decade(s) & 1951-2023$^\ast$ & 4282 & 1691 & 113\\
Fundamental problem & 1912-2023 & 2588 & 1973 & 124\\
Fundamental questions & 1973-2023 & 1589 & 414 & 70 \\
Last decade + question & 1974-2023 & 120 & 84 & 13\\ 
\hline
\multicolumn{3}{l}{$^\ast$: plus one double entry in year 1900.}\\
\end{tabular}
\caption{Expressions searched in astronomy abstracts and ADS results: covered time range and recurrence as of March 2023.}
\label{tab:incidence}
\end{table*}

\begin{figure*}[!ht]
\centering
\includegraphics[width=0.43\textwidth]{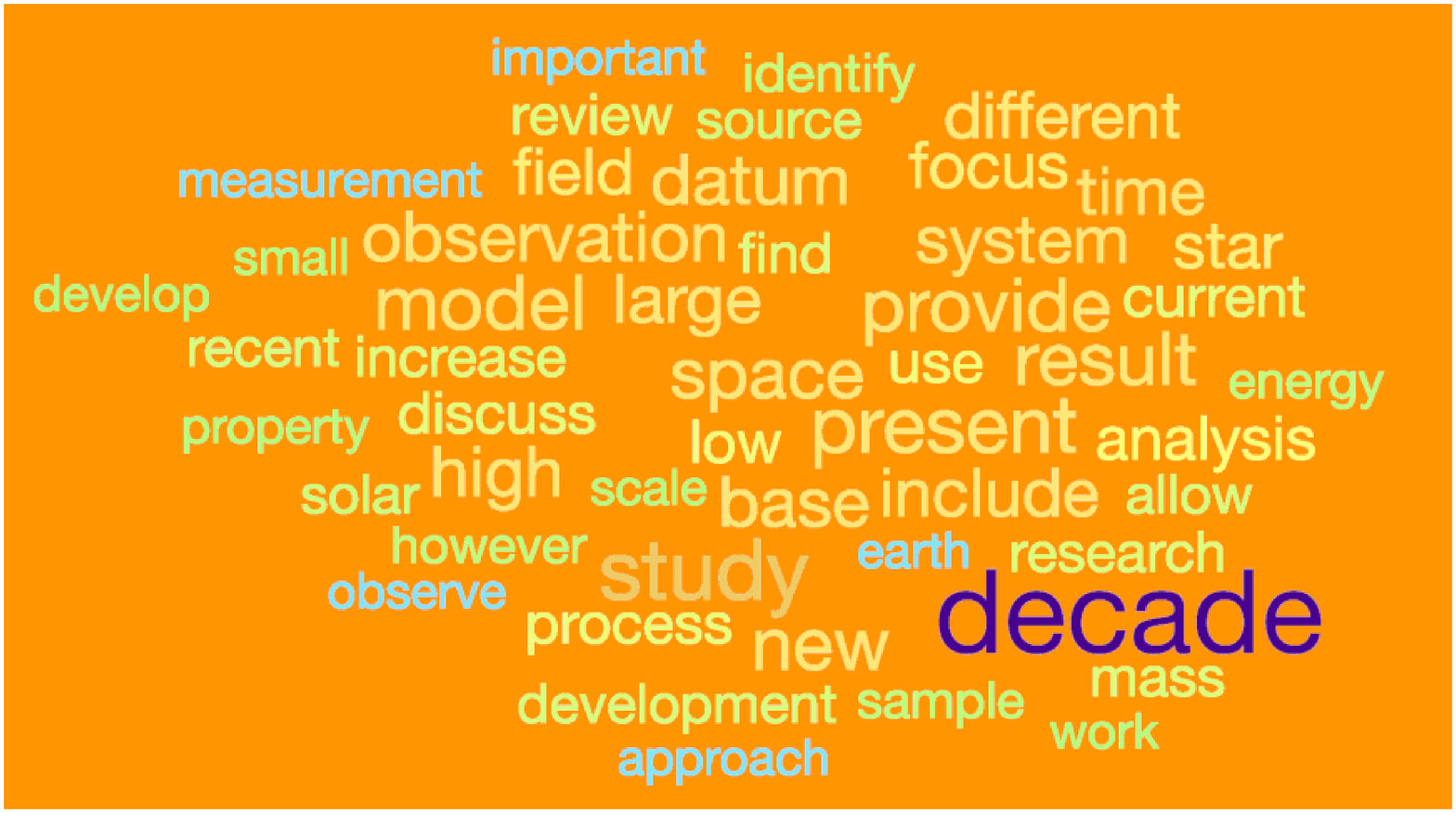}
\includegraphics[width=0.43\textwidth]{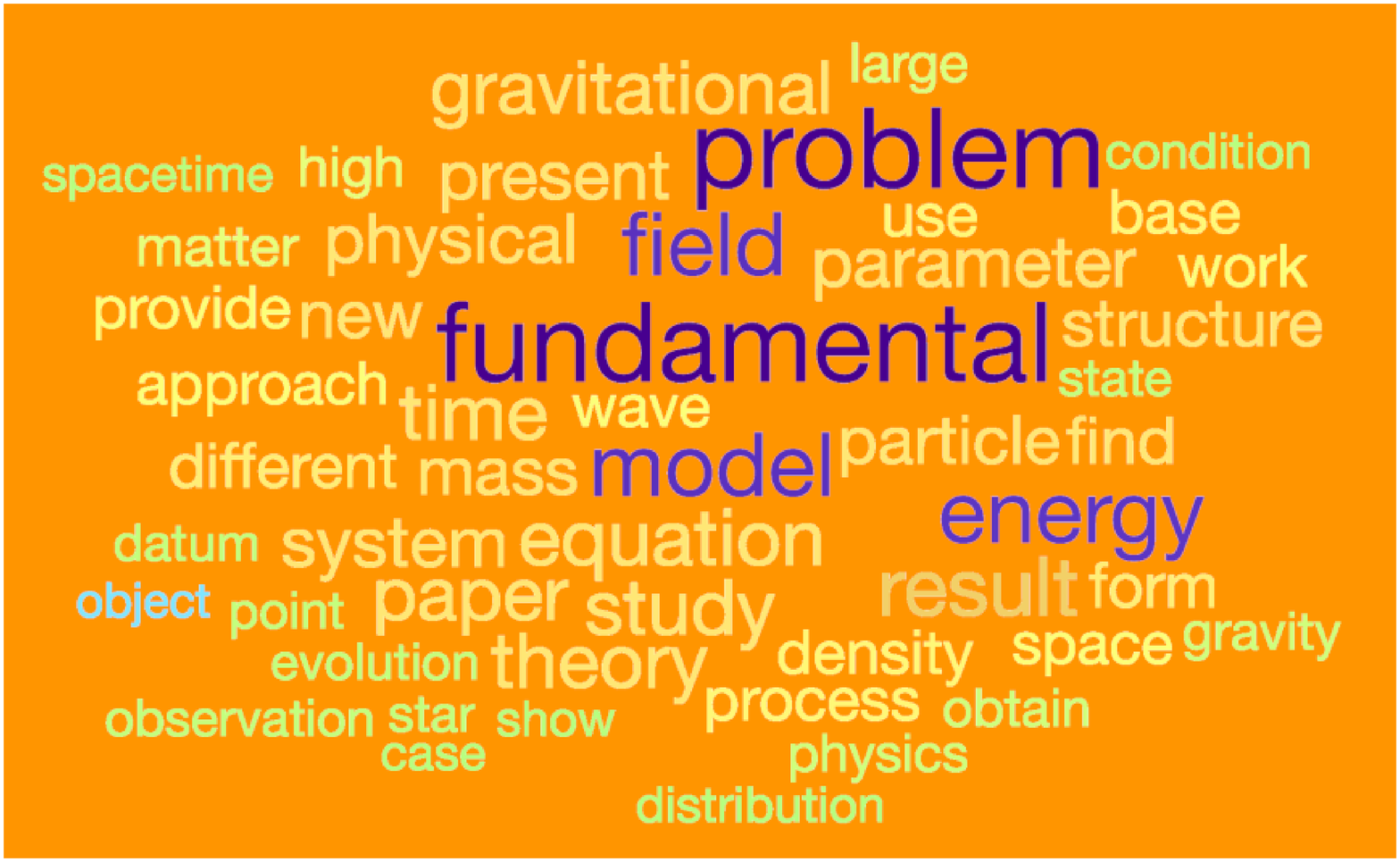}
\includegraphics[width=0.43\textwidth]{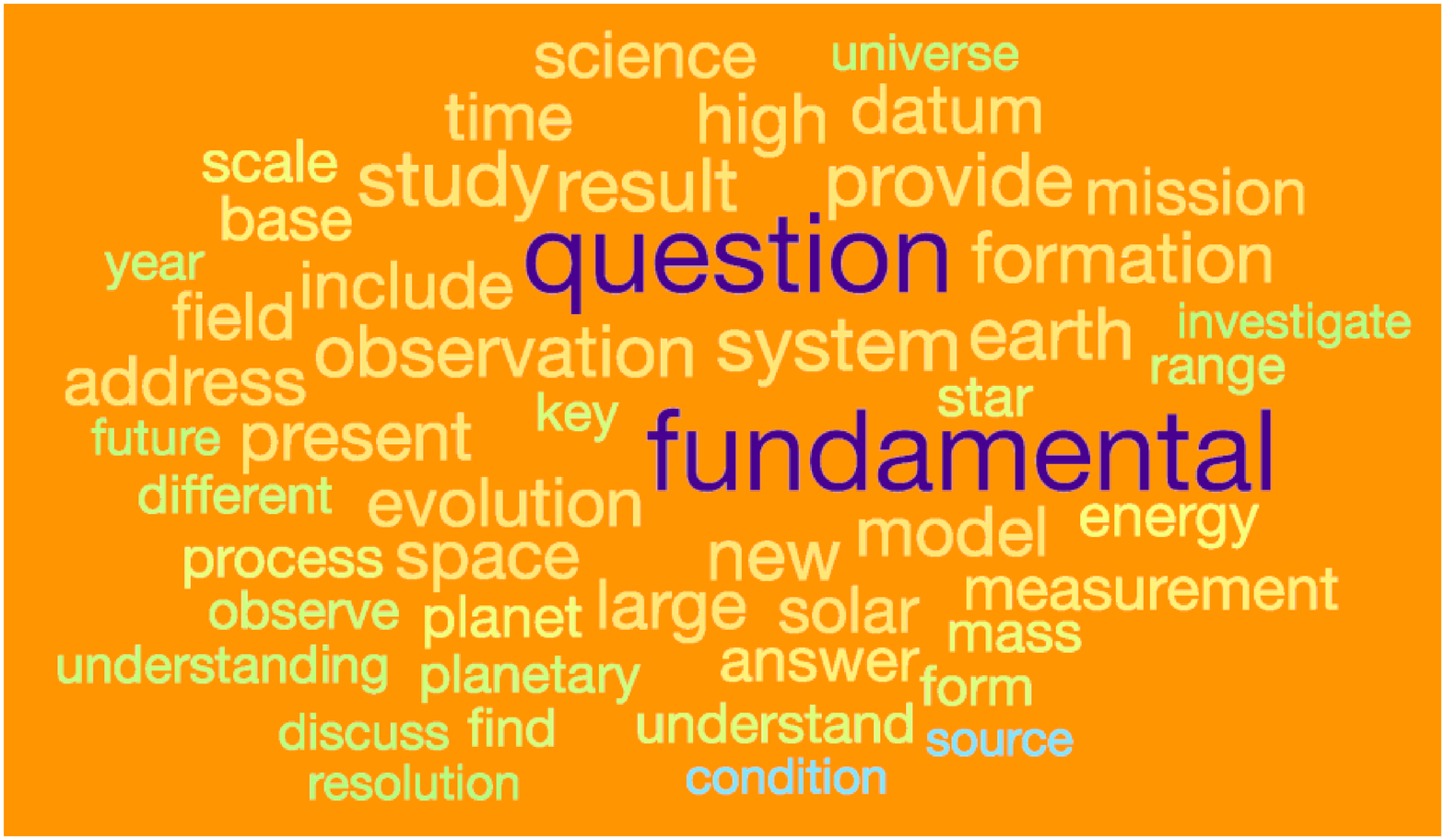}
\caption{Astronomy concept clouds since 1950 (limited to the publications shown in Fig. \ref{fig:metrics}).}
\label{fig:cloud}
\end{figure*}

\section{Current data and analysis}

We used the ADS Bibliographic Service to search in the published abstracts the recurrence of the famous ``fundamental questions'' in the ``last decades'', and some variations. Table \ref{tab:incidence} list the combinations of words we looked for.

The same Table reports also the range of years over which the given expression was actively used, the number of entries retrieved by the search engine, and the corresponding $h$-index. For convenience sake, we sort the entries by popularity.  Figure \ref{fig:metrics} shows the differential and cumulative distribution of the relevant publications through the years, from 1950 to today.

The last decades are covered by the largest number of publications. Interestingly, though, the highest ratio of refereed to non-refereed publications is found for the fundamental problems. These problems also dominate the $h$-index calculation, with an astonishing 124 (Spider), to be compared to the more humble $h=70$ of their questions cousins. 

Figure \ref{fig:metrics} also reveals that, despite being really fundamental, the problems we have been facing in the last decades have become less compelling, as demonstrated by the drop of their differential distribution since $\sim2015$. The fundamental questions, instead, become more and more pressing every day and the interest demonstrated by authors consequently rises continuously and almost monotonically.
\vspace{1.5truecm}


\section{Results}

In order to build an even stronger and more solid confirmation of our findings, we make use of one of the most fundamental statistical methods developed in the last few decades: the Concept Cloud. Figure \ref{fig:cloud} draws the result for our three main searches, and removes -- once and for all -- any possible doubt: {\it the fundamental problem of the decade is that we've been asking too many questions}. 

Keeping in mind this eye-opening evidence, we can therefore finally make a firm statement. The most popular sentence in the history of astronomy abstracts, turns out to be different, but complementary to previous statements (\citealt{torrance}; see also \citealt{king,kubrik}):\\

\noindent The most fundamental question\\
\hspace*{5pt} The most fundamental question\\
\hspace*{10pt} The most fundamental question\\
\hspace*{15pt} The most fundamental question\\
\hspace*{20pt} The most fundamental question\\
\hspace*{25pt} The most fundamental question\\
\hspace*{30pt} The most fundamental question\\
\hspace*{35pt} The most fundamental question\\
\hspace*{40pt} The most fundamental question\\
\hspace*{45pt} The most fundamental question\\
\hspace*{50pt} The most fundamental question\\
\hspace*{55pt} The most fundamental question\\
\hspace*{60pt} The most fundamental question\\
\hspace*{65pt} The most fundamental question\\
\hspace*{70pt} The most fundamental question\\
\hspace*{65pt} The most fundamental question\\
\hspace*{60pt} The most fundamental question\\
\hspace*{55pt} The most fundamental question\\
\hspace*{50pt} The most fundamental question\\
\hspace*{45pt} The most fundamental question\\
\hspace*{40pt} The most fundamental question\\
\hspace*{35pt} The most fundamental question\\
\hspace*{30pt} The most fundamental question\\
\hspace*{25pt} The most fundamental question\\
\hspace*{20pt} The most fundamental question\\
\hspace*{15pt} The most fundamental question\\
\hspace*{10pt} The most fundamental question\\
\hspace*{5pt} The most fundamental question\\
The most fundamental question\\
\noindent of all times\\
\hspace*{5pt} of all times\\
\hspace*{10pt}of all times\\
\hspace*{15pt} of all times\\
\hspace*{20pt} of all times\\
\hspace*{25pt} of all times\\
\hspace*{30pt} of all times\\
\hspace*{35pt} of all times\\
\hspace*{40pt} of all times\\
\hspace*{45pt} of all times\\
\hspace*{50pt} of all times\\
\hspace*{55pt} of all times\\
\hspace*{60pt} of all times\\
\hspace*{65pt} of all times\\
\hspace*{70pt} of all times\\
\hspace*{65pt} of all times\\
\hspace*{60pt} of all times\\
\hspace*{55pt} of all times\\
\hspace*{50pt} of all times\\
\hspace*{45pt} of all times\\
\hspace*{40pt} of all times\\
\hspace*{35pt} of all times\\
\hspace*{30pt} of all times\\
\hspace*{25pt} of all times\\
\hspace*{20pt} of all times\\
\hspace*{15pt} of all times\\
\hspace*{10pt} of all times\\
\hspace*{5pt}  of all times\\
of all times


\section{Conclusion}

If you kept reading all the way to here, it means that after all you also have nothing to do today. 

We confirm that, at least in the last five to seven decades, there has been a growing trend at seeking for the answers to the most fundamental, important, debated, and so on and so forth questions of astronomy, astrophysics, and the entire Universe itself. 

Disappointingly, with this work we have not been able to unveil the Question of all times (with capital letter), nor its answer, fundamental or not. First hints indicate that it might be related to magnetic fields.  More Patience is needed \citep{rose1988} to compete this analysis.

Nonetheless, an alarming reality came out: we keep writing the same sentences over and over again. 
Therefore we would like to conclude with an invitation to the authors of the next decades: if it is not possible to avoid those fundamentally empty phrases, please at least modify them and make them yours. Attract the attention of your readers with new, captivating expressions. In ten years from now, we would like to find and read a new list of hitherto unseen, sensational combinations of words.

\smallskip

{\it Post Scriptum:} Needless to say, this April 1st work means no offense or harm to the fans of the most fundamental questions of the last few decades. On the contrary! We love them! And we wish to keep seeking, with our astronomy colleagues and friends, the answers to the most intriguing and universal questions for many decades to come. Let's figure them all out!


\acknowledgements
We wish to thank our two anonymous colleagues for the inspiration. We are also grateful to  the one and only Ali Frolop for their influencing work. 
This research has made use of NASA’s Astrophysics Data System Bibliographic Services.
It was produced on 100\% recycled electrons. If you print it on paper, please plant a seed and grow some little tree with nice flowers... possibly orange. 
DMCC. 
Special thanks go to Stavroula, Dimitra and Angelos who helped with the ancient and modern greek translations. The cuneiform text is obviously fictional, but based on real syllabic phonetics.
No Funding Agency would ever support this study, and in fact it's been carried out exclusively in our free time... yes, we had nothing better to do last weekend.

\bibliographystyle{aa} 		
\bibliography{paper_fund_quest} 		



\end{document}